\documentstyle[12pt]{article}

\newcommand{\AmS}{{\protect\the\textfont2
  A\kern-.1667em\lower.5ex\hbox{M}\kern-.125emS}}
\newcommand{\beg}{\begin{equation}}
\newcommand{\ene}{\end{equation}}

\begin{document}

\setcounter{page}{1}
\begin{flushright}
ROME prep. 94/1013 \\
May 1994.
\end{flushright}
\vskip 10pt
\vskip 10pt

\centerline{\LARGE{\bf{
Polyakov Loops and Finite-Size Effects of}}}
\centerline{\LARGE{\bf{ Hadron Masses in Lattice Full QCD}}}
\vskip 0.5cm
\vskip 10pt
\vskip 10pt

\centerline{\bf{ S.~Antonelli$^1$, M.~Bellacci$^1$,
L.A.~Fern\'andez$^2$, A.~Mu\~noz-Sudupe$^2$, }}
\centerline{\bf{ J.J.~Ruiz-Lorenzo$^{1,2}$, R.~Sarno$^1$,
A.~Taranc\'on$^3$,}}
\centerline{ \bf and }
\centerline{\bf{ A.~Bartoloni$^1$, C.~Battista$^1$,
 N.~Cabibbo$^1$, S.~Cabasino$^1$,}}
\centerline{\bf{ E.~Panizzi$^1$, P.S.~Paolucci$^1$,
G.M~.~Todesco$^1$, M.~Torelli$^1$, R.~Tripiccione$^4$,
P.~Vicini$^1$.}}

\vskip 0.5cm
\centerline{$^1$ Dipartimento di Fisica, Universit\`a di Roma ``La
Sapienza",}
\centerline{ P.~A.~Moro, 00185 Roma, Italy and INFN, Sezione di Roma. }
\centerline{$^2$ Departamento de F\'{\i}sica Te\'orica, Universidad
Complutense de Madrid, }
\centerline {Ciudad Universitaria, 28040 Madrid, Spain.}
\centerline{$^3$ Departamento de F\'{\i}sica Te\'orica, Universidad
de Zaragoza, }
\centerline {Pza de San Francisco s/n, 50009 Zaragoza, Spain.}
\centerline{$^4$ INFN Sezione di Pisa, I-56100 Pisa, Italy.}

\vskip 10pt
\vskip 10pt
\vskip 10pt
\vskip 10pt

\begin{abstract}
The polarization of Polyakov type loops is responsible for the
dif\-ference between quenched and unquenched finite size effects on
the QCD mass spectrum. With a numerical simulation, using different
sea quarks boundary conditions, we show that we can align the
spa\-tial Polyakov loops in a predefined direction. Starting from
the\-se results, we propose a procedure to partially remove the
Polyakov type contributions in the meson propagators.
\end{abstract}

\vfill
\newpage

The finite extent of the lattice is an important source of systematic
errors in lattice QCD calculations. Theoretical and numerical analysis
\cite{THEO,AOKI} faced recently the problem of finite size effects in
full lattice QCD. The conclusions of these analysis are that the
behaviour of the hadronic masses as a function of the lattice size $L$
follows a power law $m_L = m_{\infty} + c L^{- \nu}$, that only
asymptotically goes to the exponential decay predicted by the effect
of virtual pions emitted from a point-like hadron \cite{LUSHER}. \par

The magnitude of finite-size effects is much smaller for quenched QCD
than for full QCD, especially below $L a \sim 1$fm: $\nu=$1-2 for the
quenched case and  $\nu=$2-3 for full QCD.\par The reason of the
difference can be understood using, for example, the hopping parameter
expansion on the valence quark mass that leads to the following
relation for the meson propagator \cite{AOKI}:

\begin{equation}
 \sum_{C} k_{val}^{l(C)} \langle W(C)\rangle + \sum_{C}
 k_{val}^{l(C)} \sigma_{val} \langle P(C) \rangle
\ene

\noindent where the sums extend over all possible closed paths $(C)$
of length $l(C)$, the $W(C)$ are the Wilson loops completely contained
into the lattice while $P(C)$ are valence quark loops wrapping around
the lattice in the spatial directions (Polyakov-type) and
$\langle\cdot\rangle$ denotes field averages; the $\sigma_{val}$
represents the spatial boundary conditions on the valence quarks:
$\sigma_{val} =+1$ for the periodic and $\sigma_{val}=-1$ for the
antiperiodic cases. \par

The averaged Polyakov loop $\langle P \rangle$ is different from zero
in full QCD, while it is zero in the confined phase of quenched QCD.
This means that the second term in eq.(1), due to the loops which go
around the lattice, gives different finite size effects between the
quenched and the full QCD value of mesonic correlations. \par

To obtain finite size effects comparable with those of the quenched
case we have to remove the Polyakov loop contributions. \par

In this paper we want to demonstrate that we can align the Polyakov
type loops in a predefined direction using sea quarks boundary
conditions. Starting from these results we propose a procedure to
remove the second term of eq.(1).\par


On a finite lattice with periodic boundary conditions on the gauge
fields there is a symmetry in the pure gauge action consisting in
multiplying all links in the $\mu$ direction at a constant
$x_\mu$--plane by the complex number $z_k$, that belongs to the centre
of the gauge group ($Z_3$)\cite{THOOFT}:

\begin{equation}
z_k = e^{i 2 \pi \nu_k /3} \ \ \ \ \ \   \nu_k = (0, 1, 2) \ .
\ene
Under it the Polyakov loops in the $\mu$ direction are not invariant,
but they transform as

\begin{equation}
 P \rightarrow z_k P
\ene

In full QCD we consider both the gauge and the fermionic action.
In the fermionic action

\begin{equation}
\begin{array}{rll}
S_{Wilson} &=&{\displaystyle
-k\sum_{x,\mu} \bigl(\bar \psi (x)(1-\gamma_{\mu})U_{\mu}(x)
\psi (x+\mu)
}
\\ \medskip
&&{\displaystyle
+\bar \psi (x) (1+\gamma_{\mu})U^{\dagger}_{\mu}(x-\mu)
\psi (x-\mu) \bigr)
}
\\ \medskip
&&{\displaystyle
+ \sum_x \bar \psi(x) \psi(x)
}
\end{array}
\end{equation}
the kinetic part is not invariant under $Z_3$, since on the boundary
it is not possible to find a gauge transformation of the fermionic
fields that cancels out the centre transformation on the gauge fields.
Thus the symmetry, that in the quenched confined case guarantees that
$\langle P\rangle=0$, is explicitly broken by the kinetic part of the
fermionic action. Because the non-invariant term is proportional to
$k$ the amplitude of this violation is more important for light sea
quarks.

It is possible to summarize what happens on the lattice boundary.
Ma\-king a double expansion in the full QCD, firstly with a strong
coupling  expansion on $\beta$ and secondly with a hopping parameter
expansion on the sea quark masses, we obtain  the 3--$d$ Potts Model
with magnetic field. The introduction of the fermionic action in QCD
is equivalent to turn on a magnetic field $h$ which breaks the $Z_3$
symmetry. To study the situation on the lattice  boundary we introduce
a simple model of a single spin $\Pi$ that can take the three possible
values:

\begin{equation}
\Pi_0 = 1, \ \ \ \ \Pi_1 = e^{i{2\pi/ 3}}, \ \ \ \ \Pi_2 = e^{-i{2\pi/ 3}}
\ene
with  a Hamiltonian
\begin{equation}
H = 
        h \Pi + h^{\dagger} \Pi^{\dagger}
\ene
which for $h\neq 0$ is not $Z_3$ invariant and
it is composed by terms
like those which break the $Z_3$ simmetry in the Potts Model
with magnetic field.

The $\Pi$ argument is related to the phase of Polyakov loop and
the value of $h$ with the
sea fermionic boundary conditions.

We summarize the interesting $h$ choices in fig. 1a-1e. We see that
with $h= +|h|$ (periodic boundary conditions on sea quarks) there are
two preferred states, while with $h= -|h|$ (antiperiodic boundary
conditions on sea quarks) only the $\Pi= 1$ state is selected. This
circumstance suggests a way to select the other two $Z_3$ states, as
we can see from fig. 1d and 1e. \par

Thus we expect that with periodic boundary conditions on sea quarks
both the $e^{i2\pi /3}$ and the $e^{-i2\pi /3}$ phases of the Polyakov
loops are preferred by the system. This means that for each gauge
configuration there will be domains in which they will point towards
either one of the two directions $e^{i2\pi /3}$ and $e^{-i2\pi /3}$.
No such structures exist with the antiperiodic boundary conditions and
moreover Polyakov loops are likely to point towards $ 1$ in the $Z_3$
space.

Moreover we can align the Polyakov loop in the $e^{i2\pi /3}$ (or
$e^{-i2\pi /3}$) in the $Z_3$ space if we choose $-e^{-i2\pi /3}$ (or
$-e^{i2\pi /3}$) boundary conditions on the sea quarks. \par

To check the foregoing suggestion we performed on APE100 a full QCD
simulation \cite{APE} with 2 flavors Wilson fermions at $\beta = 5.3$
on a $8^3 \times 32$ lattice with $k_{sea} =0.1670$. We performed two
different runs, one with periodic boundary conditions on the sea
quarks and the other with antiperiodic boundary conditions. We collect
in both cases 600 thermalization trajectories plus other 1200. On the
latter we perform, every 5 trajectories, a measurement of the spatial
Polyakov loops. We use the smearing procedure for the measurement of
Polyakov loops \cite{SMEAR} for 10 values of smearing.\par

The situation for the phases of the spatial Polyakov loops are
reported in figures 2 and 3. With antiperiodic boundary condition,
fig.2, we obtain that the phase is close to zero. Otherwise with
periodic boundary conditions, fig.3, we obtain that the phases are
spread in regions near $e^{i{2\pi / 3}}$ and $e^{-i{2\pi / 3}}$. \par

We also impose the boundary condition $-e^{-i2\pi /3}$ and $-e^{i2\pi
/3}$ on the sea quarks in a quick simulation on a $4^3 \times 6$
lattice with $\beta=3.0$ and $k_{sea}=0.1670$ and we verified that
Polyakov loops point towards the $e^{i2\pi /3}$ and $e^{-i2\pi /3}$
correspondingly, see fig. 4.

{}From eq.(1) (see ref. \cite{AOKI}), we do not expect large differences
between the meson masses calculated   either with periodic boundary
conditions on sea and valence quarks or antiperiodic boundary
conditions on sea and valence quarks, because the sign (not the
amplitude) of the second term is the same in both cases. But due to
our results on the polarization of the Polyakov loops we conclude that
the best boundary conditions are the antiperiodic, because their
Polyakov loops have lower dispersion than with the periodic ones.

Moreover the previous analysis gives us a procedure to partially
eliminate the Polyakov loops that contribute to eq.(1). In fact we can
choose the antiperiodic boundary conditions on the sea quarks. In this
way all the gauge configurations have Polyakov loops that point
towards the $1$ Z(3) state. Then we invert three times the fermionic
valence operator:  with antiperiodic, with $-e^{i2\pi /3}$ and with
$-e^{-i2\pi /3}$. The average of the meson propagators obtained with
the three inversions has a factor $\sigma_{val} =0$ in the second term
of eq.(1). We note that the required CPU time for valence fermionic
inversion is negligible compared to the CPU time needed to obtain the
full QCD configuration and that in the full QCD simulations the CPU
time scales as ${\rm Volume}^{{5 / 4}}$.\par The numerical simulations
of this work have been done with configurations obtained using 2
months of CPU time of a 128 nodes APE100 machine. \par
\vskip 10pt
\vskip 10pt
\vskip 10pt
\centerline { \bf Acknowledgement}
We thank G.~Parisi for many suggestions concerning this work and for
many discussions. We acknowledge interesting discussions with
E.~Marinari.\par We would like to thank F.~Marzano, J.~Pech,
F.~Rapuano for encouragements and support. \par
L.A.F.,A.M.,J.J.R.L. and A.T. acknowledge CICyT (Spain) for partial
finantial support. J.J.R.L. is also supported by a grant of MEC
(Spain).\par
\vskip 10pt
\vskip 10pt

\eject
\centerline {\bf FIGURE CAPTIONS }
\vskip 10pt
\vskip 10pt
\vskip 10pt
\vskip 10pt
\noindent {\bf Figure 1.} The energy levels of the model of eq.(6) for
different values of $h$.
The notation for the states of $\Pi$ is that of eq.(5).\par
\vskip 10pt

\noindent {\bf Figure 2.} Histogram of the phase of the x, y and z
components of the Polyakov loop for antiperiodic boundary conditions
on the sea quarks. Data are from trajectory 440 to trajectory 1800.
The value of smearing is 10. The lattice is $8^3 \times 32$. \par
\vskip 10pt

\noindent {\bf Figure 3.} The same of Figure 2. for periodic boundary
conditions on the sea quarks. \par
\vskip 10pt

\noindent {\bf Figure 4.}  Behavior of the phase of the average of the
three spatial Polyakov loops as a function of the HCMA trajectories.
Data are from trajectory 205 to trajectory 600. The value of smearing
is 0. The lattice is $4^3 \times 6$. In fig.4a we report the case of
periodic boundary conditions on the sea quarks; in fig.4b we report
the case of antiperiodic boundary conditions on the sea quarks; fig.4c
is obtained from the boundary condition $-e^{-i2\pi /3}$ and fig.4d
from the boundary condition $-e^{i2\pi /3}$. \par

\end{document}